\documentclass[pra,twocolumn,showpacs,preprintnumbers,amsmath,amssymb,floatfix]{revtex4}
\usepackage[utf8]{inputenc}
\usepackage{amsmath,amssymb}
\usepackage{graphicx}
\usepackage{braket}
\usepackage[hidelinks]{hyperref}
\newcommand{\doititle}[2]{\href{https://doi.org/#1}{#2}}
\newcommand{\arxivtitle}[2]{\href{https://arxiv.org/abs/#1}{#2}}

\begin{document}

\title{Rabi Profile Framework: A Simple Analytical and Graphical Toolkit for Quantum Control}
\author{
    Duje Bonacci\thanks{\href{https://orcid.org/0000-0002-8168-1367} {ORCID: 0000-0002-8168-1367}} \\
    Ministry of Science, Education and Youth, Zagreb, Croatia / University of Zagreb, Faculty of Croatian Studies, Zagreb, Croatia \\
    dbonacci@fhs.unizg.hr / dbonacci@mzom.hr}
    
\date{\today}

\begin{abstract}
    We introduce the \emph{Rabi profile}, the \emph{Rabi spectrum} and the \emph{Rabi leak} as triad of mathematically rigorous but conceptually transparent quantum-control tools that provide a framework for assessing the feasibility and selectivity of coherent state control in quantum systems with arbitrarily many levels, without recourse to numerical simulation.
    As the rigorous foundation for this framework, we present a self-contained discussion of the rotating-wave approximation (RWA) 
    and its limits, suitable for use as classroom material.
    We illustrate its practical utility by application to design of superconducting transmon qubits.
    Because it reduces complex quantum control questions to elementary frequency-domain reasoning, the Rabi profile framework is intended to serve equally as a research diagnostic and as a pedagogical bridge between introductory quantum mechanics and contemporary quantum technologies.

\end{abstract}

\maketitle

 \section{Introduction}

    Manipulation of discrete-level quantum systems---such as atoms, molecules, or engineered qubits---is central to a wide range of modern applications, including laser control of chemical reactions, atom optics, precision metrology, and quantum information processing~\cite{benetti2000,koch2007,brif2010}.  
    The fundamental mechanism enabling such control are \textit{Rabi oscillations}: periodic population transfer between two quantum states under the influence of a coherent, near-resonant external drive~\cite{rabi1937}. 

\subsubsection{Limitations of Established Approaches}

    The textbook treatment of Rabi oscillations is appealingly simple, typically relying on the \textit{rotating-wave approximation} (RWA)~\cite{demtroeder1988,cohentannoudji1992}.  
    RWA neglects rapidly oscillating terms in the Schrödinger equation, yielding a closed-form, sinusoidal solution for population transfer.
    Although highly accurate for weak, near-resonant drive in isolated two-level systems, the RWA breaks down under strong drive and in many-level systems.  
    
    Many studies have analyzed these breakdowns using advanced analytical or numerical approaches, addressing corrections such as the Bloch--Siegert shift and other strong-driving effects~\cite{shariar2002.1,pietikainen2017,saiko2008,forn-diaz2018,hausinger2010}.  
    However, these formulations are mathematically involved and often system-specific, making them unsuitable for quick conceptual diagnostics or pedagogical use.  
    
    Moreover, most analytical approaches focus exclusively on idealized two-level models~\cite{barata2000,fujii2003,zhicheng2024}, limiting their relevance for real systems such as polyatomic molecules, multilevel artificial atoms, or superconducting qubits. 
    In such systems, the presence of additional states can introduce significant population leakage via untargeted transitions even under nominally weak drive.

    Finally, Rabi oscillations and the RWA continue to serve as fertile ground for educational exploration and conceptual innovation.
    Many pedagogical studies have deepened intuition while offering accessible tools for instruction and simulation \cite{merlin2021,wang2008,downing2025,thimmel1999,hofferberth2006,irish2007,fujii2017,wang2015}.

\subsubsection{Novelty of Our Approach}

    We introduce a simple, analytical, and graphical toolkit---the \textit{Rabi profile framework}---that generalizes naturally to complex many-level quantum systems.  
    It enables quick and intuitive, but also rigorous, assessment of the feasibility of selective state control in quantum systems with arbitrary many coupled levels.
    
    Rather than tracking full time-dependent population dynamics, we focus on the analysis of \textit{maximum transition amplitude} achievable under a given drive strength and detuning.  
    This frequency-domain perspective reframes quantum control in terms of \textit{transition feasibility} rather than temporal evolution.  
    
    This framework is built around three constructs:
    \begin{itemize}
        \item The \textit{Rabi profile} is the Lorentzian-shaped dependence of the transition amplitude on the drive frequency, as predicted by standard Rabi theory for a single two-level transition.
        \item The \textit{Rabi spectrum} represents the superposition of all such profiles corresponding to different transitions driven by the same field, providing a complete frequency-domain picture of how the system responds to excitation.
        \item The \textit{Rabi leak} quantifies the unwanted population transfer that occurs at the resonance of the targeted transition due to overlap with neighboring profiles.
    \end{itemize}     

    The Rabi profile framework complements existing methods by offering a fast, equation-free diagnostic criterion that relies only on known transition parameters.  
    Its visual and intuitive nature makes it valuable for teaching, experimental diagnostics, and rapid design iteration.  
    Despite its simplicity, the framework is rigorously derived and quantitatively predictive, bridging the pedagogical and research domains.

\subsubsection{Outline of the Paper Structure}

    In Sec.~II, we derive a rigorous quantitative criterion for the validity of the RWA.
    
    In Sec.~III, we develop the Rabi profile framework.

    In the Appendix, we present an application of the framework to design of transmon qubit system.
    
\section{Rotating wave approximation in a two-level system}
\label{rwa2level}

    In introductory treatments of quantum dynamics, the limitations of the RWA and the role of counter-rotating terms are often hand-waved.
    Since Rabi framework hinges on thorough understanding of the limitations of RWA, we start by developing a detailed analysis of the validity range of RWA itself.

\subsection{Exact Equation of a Driven Two-Level Quantum System}
\label{full}

    Consider a two-level quantum system driven by an external field $\hat V(t)$. The time-dependent evolution of this system is governed by the Schr\"odinger equation:
    \begin{equation}
    \label{eq:Schrodinger-equation}
        i\hbar\frac{d}{dt}\,|\psi(t)\rangle
        = \bigl(\hat H_0 + \hat V(t)\bigr)\,|\psi(t)\rangle .
    \end{equation}
    The unperturbed Hamiltonian has eigenstates $|1\rangle, |2\rangle$ with eigenvalues $E_1, E_2$:
    \begin{equation}
        \hat H_0\,|i\rangle = E_i\,|i\rangle, \qquad i=1,2.
    \end{equation}
    Assuming $E_2>E_1$, we define the transition frequency:
    \begin{equation}
        \omega_0 := \frac{E_2}{\hbar} - \frac{E_1}{\hbar} \equiv \omega_2 - \omega_1 \;>\; 0 
    \end{equation}
    
    The spatially homogeneous monochromatic drive is:
    \begin{equation}
    \label{eq:driving_field}
        \hat V(t) = F \cos(\omega t)\,\hat\sigma ,
    \end{equation}
    with $F$ and $\omega$ the amplitude and frequency of the drive, and $\hat\sigma$ the coupling operator.

    The drive couples the two system's states via:
    \begin{equation}
        I_{ij} := \langle i|\hat\sigma|j\rangle , \qquad i,j\in\{1,2\},
    \end{equation}
    where coupling coefficients $I_{ij}$ satisfy:
    \begin{equation}
       I := I_{21}=I_{12} \neq 0, \qquad I_{11}=I_{22}=0 .
    \end{equation}
    
\subsubsection{Working in the Interaction Picture}

    The state of the system can be written as:
    \begin{equation}
    \label{eq:interaction-picture-solution}
        |\psi(t)\rangle 
            = a_1(t) \, | 1 \rangle \, e^{-i \omega_1 t}
            + a_2(t) \, | 2 \rangle \, e^{-i \omega_2 t}
    \end{equation}
    separating the unperturbed part of the evolution---complex exponential terms of constant (unit) amplitude---from the time- and drive-dependent amplitudes $a_1(t)$ and $a_2(t)$.
    
    The instantaneous population of level $i$ is given by:
    \begin{equation}
    \label{eq:population-amplitude}
        \Pi_i(t) = |a_i(t)|^2, \quad 0 < \Pi_i(t) < 1, \quad i = 1, 2,
    \end{equation}
    with normalization enforcing  $\Pi_1(t) + \Pi_2(t) = 1$.
    
    Combining Eq.~\eqref{eq:interaction-picture-solution} with a vector representation:
    \begin{equation}
        \mathbf{a}(t) := 
        \begin{bmatrix}
            a_1(t) \\
            a_2(t)
        \end{bmatrix},
    \end{equation}
    expanding the cosine term in Eq.~\eqref{eq:driving_field} using the Euler identity, and rearranging, Eq.~\eqref{eq:Schrodinger-equation} transforms into:
    \begin{widetext}
    \begin{equation}
    \label{eq:coupled-equations}
        \frac{d}{dt} \mathbf{a}(t) 
        = -i \frac{F I}{2 \hbar}
            \begin{bmatrix}
                0 & e^{i (\omega - \omega_0) t} + e^{-i (\omega + \omega_0) t} \\
                e^{-i (\omega - \omega_0) t} + e^{i (\omega + \omega_0) t} & 0
            \end{bmatrix}
        \mathbf{a}(t).
    \end{equation}
    \end{widetext}
    
    Equation~\eqref{eq:coupled-equations} is the exact dynamical equation of the two-level system, including both (co-)rotating ($\omega - \omega_0$) and counter-rotating ($\omega + \omega_0$) terms.

\subsubsection{Recasting Equations into Natural Units}

    We next introduce a set of dimensionless variables that absorb the system's parameters into the time and frequency scales themselves. 
    
    We start by defining the effective drive intensity:
    \begin{equation}
    	D := \frac{F I}{\hbar},
    \end{equation}
    which sets the absolute scale for coupling between the two levels induced by the external drive, and is identical to the on-resonance Rabi frequency $\Omega_R = FI/\hbar$ of standard treatments~\cite{cohentannoudji1992, foot2005}.
    Rather than using $\Omega_R$, we introduce the symbol $D$ to emphasize its role as the natural frequency scale of our dimensionless framework, making the structure of the subsequent analysis more transparent.
    
    We then introduce a dimensionless time variable $\tau$, and call it \textit{proper time} of the system:
    \begin{equation}
    \label{tau}
        \tau := \frac{D}{2\pi} t.
    \end{equation}
    We shall demonstrate that under resonant conditions ($\omega = \omega_0$) and within the bounds of validity of the RWA:
    \begin{itemize}
        \item $\tau = 1$ corresponds to a single full population transfer from one level to the other;
        \item $\tau = 2$ represents the duration of one complete Rabi oscillation (forward and reverse transfer);
    \end{itemize}
          
    Next, we define the normalized drive frequency $\Gamma$:
    \begin{equation}
    \label{eq:Gamma}
        \Gamma(\omega) := \frac{\omega}{D}
        \equiv \frac{\omega}{\Omega_R}.
    \end{equation}
       
    $\Gamma$ is a dimensionless scaling parameter whose role we shall show to be analogous---albeit inverse---to the conventional Rabi frequency $\Omega_R$: it serves as a "power knob" that gauges the system’s response to external drive.
   
    At resonance, when $\omega = \omega_0$, we define the corresponding constant:
    \begin{equation}
    \label{eq:Gamma_0}
        \Gamma_0 := \Gamma(\omega=\omega_0) = \frac{\omega_0}{D}.
    \end{equation}

    We finally define the normalized detuning of the drive frequency from the system-transition resonant one:
    \begin{equation}
    \label{eq:Delta}
        \Delta(\omega) 
            := \frac{\omega - \omega_0}{D} 
            = \Gamma(\omega) - \Gamma_0.
    \end{equation}
    
    In terms of the normalized variables, Eq.~\eqref{eq:coupled-equations} becomes:
    \begin{widetext}
    \begin{equation}
    \label{eq:matrix-form}
        \frac{d}{d \tau} \mathbf{a}(\tau)
        = -i \pi
        \begin{bmatrix}
        0 & e^{i 2\pi \Delta \tau} \big(1 + e^{-i 4 \pi \Gamma \tau} \big) \\
        e^{-i 2\pi \Delta \tau} \big(1 + e^{i 4 \pi \Gamma \tau} \big) & 0
        \end{bmatrix}
        \mathbf{a}(\tau).
    \end{equation}
    \end{widetext}

\subsubsection{Decoupling Dynamical Effects of Detuning and Counter-Rotation}
\label{ss:decoupling}

    Lastly, we introduce one final substitution:
    \begin{equation}
    \label{eq:transformation}
        \mathbf{a}(\tau) 
        = 
        \begin{bmatrix}
            e^{i \pi \Delta \tau} & 0 \\
            0 & e^{-i \pi \Delta \tau}
        \end{bmatrix}
        \mathbf{\tilde{a}}(\tau),
    \end{equation}
    which transforms Eq.~\eqref{eq:matrix-form} into:
    \begin{equation}
    \label{eq:final-full-equation}
        \frac{d}{d\tau} \mathbf{\tilde a}(\tau) 
        = -i \pi 
            \begin{bmatrix}
                \Delta & 1 + e^{-i 4 \pi \Gamma\tau} \\ 
                1 + e^{i 4 \pi \Gamma\tau} & -\Delta
            \end{bmatrix}
            \mathbf{\tilde a}(\tau).
    \end{equation}
    
    Eq.~\eqref{eq:final-full-equation} is still \textit{exact}: it captures the full system dynamics across all regimes of external drive, whether weak or strong, resonant or not. 
           
    Importantly, this form reveals a key structural insight: the detuning---quantified by $ \Delta $, and the counter-rotation---governed by $\Gamma$, act as mutually independent generators of the system's dynamics.
    This decoupling immediately leads to two significant insights. 
    
    First, it allows for a rough quantitative criterion for the validity of RWA. The approximation holds when:
    \begin{equation}
    \label{eq:RWA-validity-criterion-rough}
        \Gamma \equiv \frac{\omega}{D} \gg 1,
    \end{equation}
    i.e., when the effective drive intensity is much weaker than the energy scale set by the external drive frequency. When this condition holds, oscillations of the off-diagonal terms are so rapid that they cancel out at a scale well below a single unit of proper time, and their contribution to the dynamics of the system can thus be negledted.

    Second, it clarifies the dynamical role of detuning.
    The diagonal terms $\Delta$ do not generate the population transfer, but---on contrary---shield a share of the population from participating in that transfer.
    
\subsection{RWA Solution and Beyond}

    We shall now explore the solutions of Eq.~\eqref{eq:final-full-equation} in different dynamical regimes.
    
\subsubsection{Traditional Rotating Wave Approximation Solution}

    First we explore the traditional RWA solution. 

    The fundamental assumption of the RWA is that the counter-rotating elements provide negligible contribution to the system's total dynamics, and can thus be neglected. 
    Applying RWA to Eq.~\eqref{eq:final-full-equation} yields:
    \begin{equation}
    \label{eq:RWA}
        \frac{d}{d \tau} \mathbf{\tilde a}(\tau)
        \overset{\scriptscriptstyle\mathrm{RWA}}{\approx} -i \pi
            \begin{bmatrix}
                \Delta & 1 \\ 
                1 & -\Delta
            \end{bmatrix}
            \mathbf{\tilde a}(\tau).
    \end{equation}
    
    Differentiating again with respect to $\tau$, we find:
    \begin{equation}
    \label{eq:RWA_final}
        \frac{d^2}{d \tau^2} \mathbf{\tilde{a}}(\tau)
        \overset{\scriptscriptstyle\mathrm{RWA}}{\approx} 
        - \pi^2 \big(1 + \Delta^2\big)
        \begin{bmatrix}
            1 & 0 \\
            0 & 1
        \end{bmatrix}
        \mathbf{\tilde{a}}(\tau),
    \end{equation}
    which describes decoupled simple harmonic oscillations of the amplitudes of two system levels.
    
    Solving Eq.~\eqref{eq:RWA_final} yields the textbook RWA solution. The half-period $T_\frac{1}{2}$ of population oscillations between the two levels (in units of dimensionless time $\tau$) and the corresponding oscillation amplitude $A$ (between 1 - full transfer, and 0 - no transfer) depend solely on the normalized detuning $\Delta$:
    \begin{equation}
    \label{eq:RWA_X_Rabi}\  
        A(\Delta) := \frac{1}{1 +  \Delta^2},
        \qquad
        T_{\frac{1}{2}}(\Delta) := \frac{1}{\sqrt{1 + \Delta^2}}.
    \end{equation}
       
    Exactly at resonance ($\Delta = 0$), we get the complete population inversion for each unit of proper time:
    \begin{equation}
        A_0 := A(\Delta = 0) = 1,
        \qquad
        T_{\frac{1}{2}}^0 := T_{\frac{1}{2}}(\Delta = 0) = 1.
    \end{equation}

\subsubsection{Correction Due to the Counter-Rotating Terms}
    
    Next we calculate the correction to the RWA solution by expanding the full solution perturbatively as:
    \begin{equation}
    \label{eq:non-RWA-correction}
        \mathbf{\tilde{a}}(\tau) = \mathbf{\tilde{a}}^{\scriptscriptstyle{\text{RWA}}}(\tau) + \mathbf{\tilde{a}}^\text{per}(\tau), \quad
        |\mathbf{\tilde{a}}^\text{per}|^2 \ll |\mathbf{\tilde{a}}^{\scriptscriptstyle{\text{RWA}}}|^2
    \end{equation}
    where $\mathbf{\tilde{a}}^{\scriptscriptstyle{\text{RWA}}}(\tau)$ is the RWA solution, and $\mathbf{\tilde{a}}^\text{per}(\tau)$ is the perturbative correction.
 
    Introducing expansion Eq.~\eqref{eq:non-RWA-correction} into Eq.~\eqref{eq:final-full-equation} and rearranging, we obtain: 
    \begin{widetext}
    \begin{equation}
    \label{eq:full-B-S-equation}
        \frac{d}{d\tau} \mathbf{\tilde a}^\text{per}(\tau) 
            = -i \pi \left(
            \begin{bmatrix}
                \Delta & 1 +e^{-i 4 \pi \Gamma \tau} \\
                1 +e^{i 4 \pi \Gamma \tau} & -\Delta
            \end{bmatrix}
            \mathbf{\tilde a}^\text{per}(\tau)
            +
            \begin{bmatrix}
                0 &  e^{-i 4 \pi \Gamma \tau} \\
                e^{i 4 \pi \Gamma \tau} & 0
            \end{bmatrix}
            \mathbf{\tilde a}^{\scriptscriptstyle{\text{RWA}}}(\tau)
            \right)
    \end{equation}
    \end{widetext}

    If the condition Eq.~\eqref{eq:non-RWA-correction} is satisfied, then Eq.~\eqref{eq:full-B-S-equation} can be approximated by:
    \begin{equation}
    \label{eq:non-RWA-correction-equation}
        \frac{d}{d\tau} \mathbf{\tilde a}^\text{per}(\tau) 
             \overset{\text{per}}{\approx}
            -i \pi
            \begin{bmatrix}
                0 & e^{-i 4 \pi \Gamma \tau} \\
                e^{i 4 \pi \Gamma \tau} & 0
            \end{bmatrix}
            \mathbf{\tilde a}^{\scriptscriptstyle{\text{RWA}}}(\tau)
    \end{equation}
    
    Thus, a formal solution for $\mathbf{\tilde a}^{\text{per}}$ can be obtained by integrating with respect to $\tau$:    
    \begin{equation}
    \label{eq:non-RWA-correction-solution}
        \mathbf{\tilde a}^\text{per}(\tau') 
             \overset{\text{per}}{\approx}
            -i \pi \int_{\tau_0}^{\tau'} d\tau
            \begin{bmatrix}
                0 & e^{-i 4 \pi \Gamma \tau} \\
                e^{i 4 \pi \Gamma \tau} & 0
            \end{bmatrix}
            \mathbf{\tilde a}^{\scriptscriptstyle{\text{RWA}}}(\tau)
    \end{equation}

    This integral can be approximately evaluated if frequency of the (counter-)rotation of matrix elements and frequency of the population oscillations predicted by the RWA solution significantly differ.
    
    To proceed with integration, consider the effective number of full cycles that the exponential counter-rotating terms $e^{\pm i 4 \pi \Gamma\tau}$ perform during one half-cycle $T_\frac{1}{2}$ of the Rabi oscillations predicted by the RWA solution:
    \begin{equation}
    \label{eq:N-of-rounds}
        N = 2 \, \Gamma \, T_{\frac{1}{2}}. 
    \end{equation}
    For $N \gg 1$ the matrix elements rotate much faster, and for $N \ll 1$ the roles reverse---the RWA solution rotates much faster.
    
    In both cases, we can treat the slower element as an adiabatic parameter and pull it outside the integral.
    
    \paragraph{Case 1: $N \ll 1$.}
    When $N \ll 1$, the matrix elements rotate much slower than the RWA solution and can thus be considered as adiabatic parameters. Eq.~\eqref{eq:non-RWA-correction-solution} can be approximated with:    
    \begin{equation}
    \label{eq:non-RWA-correction-result-2}
        \mathbf{\tilde a}^\text{per}(\tau') 
            \underset{N \ll 1}{\overset{\text{per}}{\approx}}
            - i \pi
             \begin{bmatrix}
                0 & e^{-i 4 \pi \Gamma \tau_0} \\
                e^{i 4 \pi \Gamma \tau_0} & 0
            \end{bmatrix}
             \int_{\tau_0}^{\tau'} d\tau \, \mathbf{\tilde a}^{\scriptscriptstyle{\text{RWA}}}(\tau)
    \end{equation}
    
    The remaining integral simply yields a phase-offset version of the RWA solution, while the prefactor enhances its amplitude by a factor of the order of $\pi$. Consequently, $| \mathbf{\tilde a}^\text{per} |^2$ can become comparable to or even larger than $| \mathbf{\tilde a}^{\scriptscriptstyle{\text{RWA}}} |^2$.
    This shows that the criterion of Eq.~\eqref{eq:non-RWA-correction} is violated, and the RWA solution does not provide a reliable description of the system's evolution in this regime.
    
    Intuitively, what happens is that $N\ll 1$ corresponds to an extremely intense drive.
    Due to its intensity, such drive induces many population transfers within every single half-cycle of the drive's own oscillations.
    
    \paragraph{Case 2: $N \gg 1$.} 
    If $N \gg 1$---i.e. if the exponential terms rotate with a frequency much higher than that of the RWA solution, Eq.~\eqref{eq:RWA_X_Rabi}---then 
    $\mathbf{\tilde a}^{\scriptscriptstyle{\text{RWA}}}(\tau)$ can be treated as an adiabatic parameter and extracted outside the integral:
    \begin{equation}
    \label{eq:non-RWA-correction-adiabatic}
        \mathbf{\tilde a}^\text{per}(\tau') 
             \underset{N \gg 1}{\overset{\text{per}}{\approx}}
            -i \pi 
          \left(  \int_{\tau_0}^{\tau'} d\tau
            \begin{bmatrix}
                0 & e^{-i 4 \pi \Gamma \tau} \\
                e^{i 4 \pi \Gamma \tau} & 0
            \end{bmatrix}            
            \right)\, 
            \mathbf{\tilde a}^{\scriptscriptstyle{\text{RWA}}}(\tau_0)
    \end{equation}
    which can be integrated to yield
    \begin{equation}
    \label{eq:non-RWA-correction-result}
        \mathbf{\tilde a}^\text{per}(\tau') 
             \underset{N \gg 1}{\overset{\text{per}}{\approx}}
           \left( \left .
             \begin{bmatrix}
                0 & e^{-i 4 \pi \Gamma \tau} \\
                -e^{i 4 \pi \Gamma \tau} & 0
            \end{bmatrix}
            \right |_{\tau=\tau_0}^{\tau=\tau'} \right)
            \frac{\mathbf{\tilde a}^{\scriptscriptstyle{\text{RWA}}}(\tau_0)}{4 \, \Gamma} .
    \end{equation}
    
    These corrections appear as small, rapid oscillations of frequency $2\Gamma$ superimposed on the slower Rabi envelope, known as \emph{Bloch--Siegert oscillations}.~\cite{bloch1940}.
    
    As drive intensity increases, the amplitude and frequency of the Bloch-Siegert oscillations become comparable to those of the RWA solution.
    This leads to small, yet noticeable deviations from the pure population oscillations of pure RWA regime.
    This effect is clearly observable in Fig.~\ref{fig:TwoLevel_Multi} (middle column).
    
\subsubsection{Establishing the RWA Solution Validity Criteria}

    Since the amplitude of the terms in the matrix in Eq.~\eqref{eq:non-RWA-correction-result} is $1$, the ratio of amplitudes of the RWA and perturbative solution at any time during the system's evolution satisfies:
    \begin{equation}
        \frac{| \mathbf{\tilde a}^\text{per} |^2 }{| \mathbf{\tilde a}^{\scriptscriptstyle{\text{RWA}}} |^2}
         \underset{N \gg 1}{\overset{\text{per}}{\approx}}
         \frac{1}{{16 \, \Gamma^2} }
    \end{equation}

    The condition of Eq.~\eqref{eq:non-RWA-correction} is thus fulfilled---and the RWA solution is a fair description of system's evolution---as long as:
    \begin{equation}
    \label{eq:RWA-validity-criterion-1}
        \Gamma \equiv \frac{\omega}{D} \gtrsim 1 
        \quad \Rightarrow \quad
        \omega \gtrsim \, D
    \end{equation}

    However, in order to fully verify the validity of RWA solution in this case, we also have to examine what limitations the necessary precondition $N \gg 1$ imposes.
    Replacing $T_{\frac{1}{2}}$ in Eq.~\eqref{eq:N-of-rounds} with its definition, Eq.~\eqref{eq:RWA_X_Rabi} and rearranging, $N \gg 1$ can be expressed as:
    \begin{equation}
        \underbrace{\Gamma}_\text{lhs} \gg \frac{1}{2}\underbrace{\sqrt{1+\Delta^2}}_\text{rhs}
    \end{equation}
    Employing Eq.~\eqref{eq:Gamma} and Eq.~\eqref{eq:Delta}, squaring both sides, approximating $lhs^2 \gg \frac{1}{4} rhs^2$ with $lhs^2 \gtrsim rhs^2$ and rearranging leads to:
    \begin{align}
        \omega \gtrsim \frac{\omega_0}{2}\left( 1 + \frac{1}{\Gamma_0^2} \right).
        \label{eq:RWA-robust-threshold}
    \end{align}
    Combined with Eq.~\eqref{eq:Gamma_0} and Eq.~\eqref{eq:RWA-validity-criterion-1}, this brings us to the most complete version of the RWA solution validity criterion:
    \begin{equation}
        \omega \gtrsim \max \left\{
            D, \,
            \frac{\omega_0}{2} \left ( 1 + \left( \frac{D}{\omega_0} \right)^2 \right)
        \right\}.
        \label{eq:RWA-composite}
    \end{equation}

    This composite condition unifies two complementary requirements on the frequency of the external drive:
    \begin{itemize}
        \item in order for the counter-rotating terms to be adiabatically decoupled from the RWA solution, the drive frequency must exceed the threshold set by the system’s resonant frequency and the effective-drive-intensity-to-resonant-frequency ratio;
        \item in order to suppress perturbations from Bloch--Siegert oscillations, it must also exceed the limit set by effective drive intensity.
    \end{itemize}
    Together, these conditions delineate the precise parameter region where the RWA solution is robust.
    
    We can cast Eq.~\eqref{eq:RWA-composite} in an even more compact form by introducing two dimensionless variables:
    \begin{equation}
    \label{eq:rho-and-R}
        \rho := \frac{\omega}{\omega_0}, 
        \qquad 
         R := \frac{1}{\Gamma_0} = \frac{D}{\omega_0}.
    \end{equation}
    Here $\rho$ is the relative drive frequency normalized by the transition frequency, while $R$---the inverse of $\Gamma_0$---is the analogously normalized effective drive intensity.
    
\begin{figure}
        \centering
        \includegraphics[width=\columnwidth]{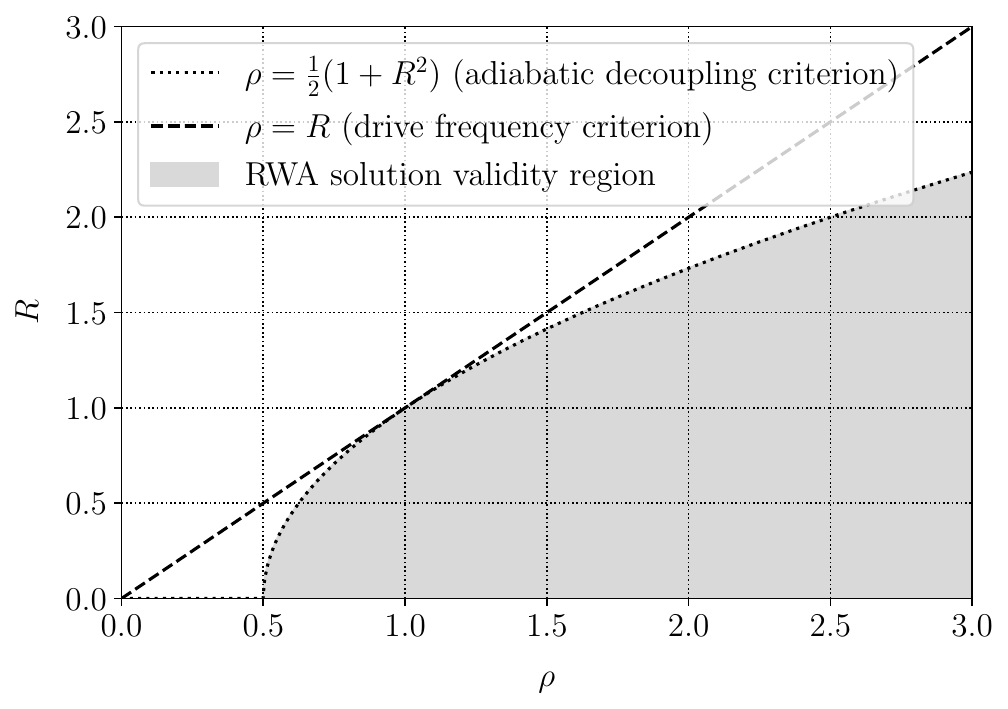}
        \caption{Visual representation of the composite validity criteria of RWA solution to the two-level system dynamics.}
    \label{fig:RWA-validity-range}
    \end{figure}

    In these variables, the composite validity condition becomes:
    \begin{equation}
        \rho \gtrsim \max \!\left\{ R, \; \tfrac{1}{2}\left(1+R^2\right) \right\}.
    \label{eq:RWA-criteria-rhoR}
    \end{equation}
    Figure ~\ref{fig:RWA-validity-range} provides a visual presentation of this relation.
    
    The first thing this form makes transparent is that the adiabatic decoupling criterion is actually more restrictive of the two, and if it is fulfilled, then the drive frequency one is automatically so too.  
    
    Second, it is evident that if $R \ll 1$ both criteria are satisfied and the RWA remains valid for any drive frequency greater than half the transition frequency.
    Granted $R \ll 1$ holds, the RWA solution faithfully captures the dynamics across a broad range of frequencies well below, at, and anywhere above the system's resonance.

    We thus obtain a simple and general criterion for the validity of the RWA solution, which is a mathematically justified refinement of Eq~\eqref{eq:RWA-validity-criterion-rough}: 
    \begin{equation}
    \label{eq:simplified-RWA-validity-criterion}
        R \ll 1 \quad \Leftrightarrow \quad \Gamma_0 \gg 1    
    \end{equation}
    This result is the foundation of Rabi profile framework.
    
\subsubsection{Extremely-Driven System Dynamics and the Non-RWA Solution}

    For completeness, we conclude the analysis of the dynamics of the externally driven two-level system with the case of extremely strong drive (XD), characterized by $\Gamma \ll 1$, which is rarely discussed in textbooks. 
    
    In this regime, the population oscillates much more rapidly than the drive.
    Thus, we can consider the counter-rotating terms in Eq.~\eqref{eq:final-full-equation} to be adiabatic, so that the equation transforms into:
    \begin{equation}
    \label{eq:extreme-drive-equation}
        \frac{d}{d\tau} \mathbf{\tilde a}(\tau) 
        \overset{\scriptscriptstyle\mathrm{XD}}{\approx} -i \pi 
            \begin{bmatrix}
                \Delta & 1 + e^{-i \Phi} \\ 
                1 + e^{i \Phi} & -\Delta
            \end{bmatrix}
            \mathbf{\tilde a}(\tau),
    \end{equation}
    where $\Phi$ is an adiabatically varying phase. 
    
    We first rewrite the off-diagonal terms as:
    \begin{equation}
        1+e^{\pm i \Phi} = 2 e^{\pm i\Phi/2} \cos(\Phi/2).       
    \end{equation}
    then, introducing the unitary transformation:
    \begin{equation}
       \mathbf{\tilde a}(\tau) = 
        \begin{bmatrix}
            e^{-i\Phi/4} & 0 \\
            0 & e^{+i\Phi/4}
        \end{bmatrix}
        \mathbf{b}(\tau),
    \end{equation}
    Eq.~\eqref{eq:extreme-drive-equation} becomes:
    \begin{equation}
    \frac{d}{d\tau}\mathbf b(\tau) 
    \overset{\scriptscriptstyle\mathrm{XD}}{\approx}
    -i\pi
    \begin{bmatrix}
    \Delta & g \\ g & -\Delta
    \end{bmatrix}\mathbf b(\tau),
    \quad g := 2\cos\!\frac{\Phi}{2},
    \end{equation}
    which finally leads to the Rabi-like population oscillations with the following parameters:   
    \begin{equation}
        \label{eq:XD-solution}
        T_\frac{1}{2}^{\mathrm{XD}}(g, \Delta) 
        = \frac{1}{\sqrt{ g^2 + \Delta^2}} , \quad
        A^{\mathrm{XD}}(g, \Delta) = \frac{g^2}{g^2 + \Delta^2} 
    \end{equation}
    
    At resonance ($\Delta=0$) the system again goes through complete population inversions ($A^{\mathrm{XD}} = 1$), but now with a sinusoidally time-varying transfer period.
    At $\Phi=\pi$ (mod $2\pi$), $g=0$ and population oscillations instantaneously come to a halt and the direction of transfer reverses.
    These effects can be observed in the numerical simulations presented in Fig.~\ref{fig:TwoLevel_Multi} (bottom right plot).

\section{Rabi Profile Framework}
\label{graphical analysis}    

    In this section, we introduce the concept of the \textit{Rabi profile}---the central element of our approach.  
    
    Two complementary quantities naturally follow: the \textit{Rabi spectrum}, describing the combined frequency-domain response of all coupled transitions, and the \textit{Rabi leak}, a quantitative measure of unwanted off-resonant excitation. Together, they form a unified framework for analyzing the controlled manipulation of arbitrarily complex quantum systems.
    
    \subsection{Rabi Profile of a Single Transition}
    \label{single transition}
    
    For a single transition between two states of a quantum system, the \textit{Rabi profile} $P$ is defined as the amplitude of population oscillations predicted by the RWA solution, Eq.~\eqref{eq:RWA_X_Rabi}, expressed as a function of the drive amplitude $F$ and frequency $\omega$ (Eq.~\eqref{eq:driving_field}):
    \begin{equation}
    \label{eq:Rabi-profile-1} 
        P(F, \omega)
        := A(\Delta(F, \omega))
        = \frac{1}{1 +  \left(\frac{\hbar(\omega - \omega_0)}{F I}\right)^2}.
    \end{equation}
    When the detuning parameter $\Delta$ (Eq.~\eqref{eq:Delta}) is written in normalized form using $\rho$ and $R$ (Eq.~\eqref{eq:rho-and-R}), the same function can be expressed as
    \begin{equation}
    \label{eq:Rabi-profile}
        \Delta = \frac{\rho - 1}{R}
        \ \Rightarrow \
        P(R, \rho) = \frac{1}{1+\left(\frac{\rho - 1}{R}\right)^2}.
    \end{equation}
    
    The Rabi profile is thus a Lorentzian function centered at $\rho=1$ (or $\omega=\omega_0$) with a half-width at half-maximum equal to $R$ (or $F I/\hbar \equiv D$ in frequency units).  
    Analytically, it encodes the dependence of transition amplitude on detuning and drive strength;  
    graphically, it provides a direct visual criterion for the validity of the RWA condition, Eq.~(\ref{eq:simplified-RWA-validity-criterion}):  
    the RWA holds when the profile remains narrow ($R \ll 1$) so that its value at $\rho=0$ is negligible.  

\subsubsection{Resonant Drive Dynamics}    
    
    Figure~\ref{fig:RabiProfile_Multi} presents Rabi profiles corresponding to several different values of normalized effective drive intensity $R$.

    For weak $R$ (narrow Rabi profiles), the validity range of the RWA solution includes the resonance of the system.
    As $R$ increases, the profile broadens and the range of drive frequencies for which the RWA solution is valid shifts to higher values.
    Beyond $R \approx 1$, RWA solution is no longer valid at or near---but only far above---the resonance.

    Figure~\ref{fig:TwoLevel_Multi} shows numerical solutions to the full system dynamics for a resonantly driven two-level system, corresponding to the Rabi profiles from Fig.~\ref{fig:RabiProfile_Multi}.
    For narrow profiles, the actual system's dynamics closely matches the RWA prediction (left column plots)---pure sinusoidal population oscillations.
    As the drive intensity increases and the value for Rabi profile at the origin becomes non-negligible---albeit still rather small---the Bloch-Siegert oscillations become noticeable (middle column plots).
    With an additional increase in the drive strength, the RWA solution becomes obsolete, and eventually the extreme drive solution, Eq.~\eqref{eq:XD-solution}, takes over (right column plots).
    
    \begin{widetext}
    
    \begin{figure}
        \centering
        \includegraphics[width=\textwidth]{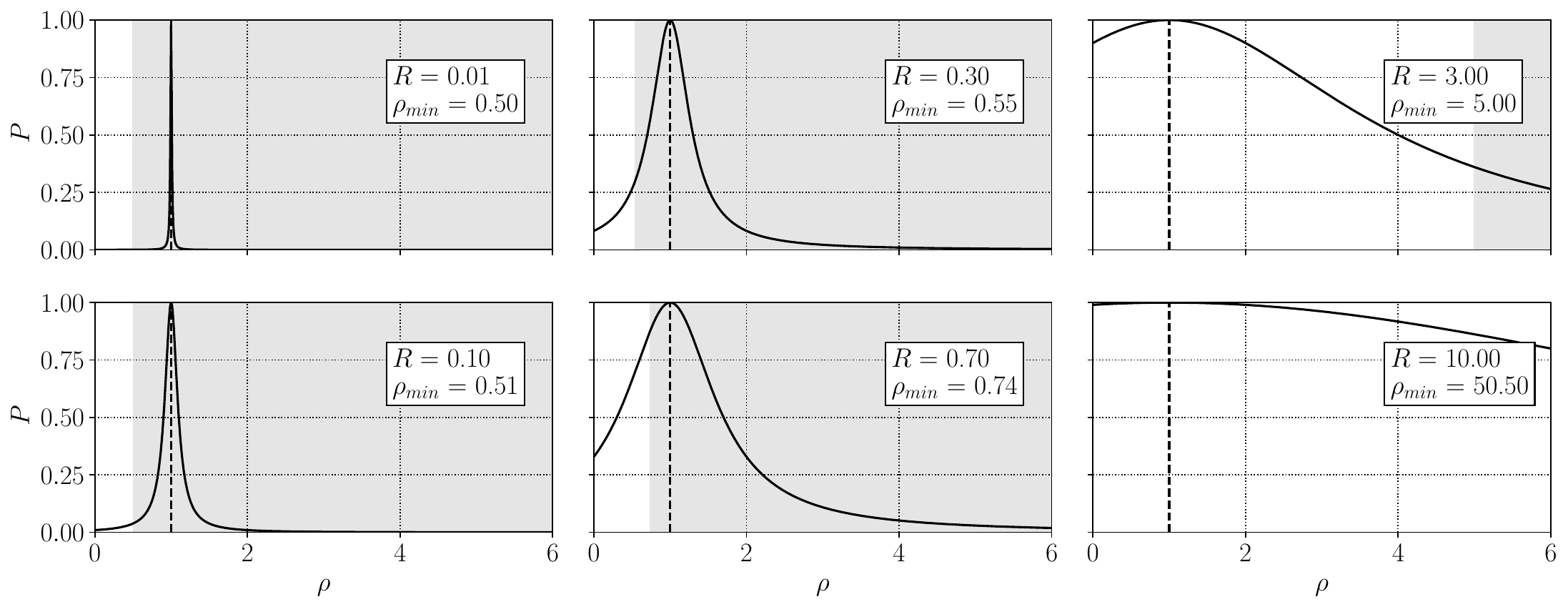}
        \caption{Rabi profiles for a single transition for several different values of normalized effective drive intensity $R$.
        Shaded area indicates the region of relative drive frequency $\rho > \rho_{min} := \tfrac{1}{2}(1+R^2)$ where the RWA solution is valid.
        The vertical dashed line indicates the resonant frequency ($\rho = 1$).}
        \label{fig:RabiProfile_Multi}
    \end{figure}
    
    \begin{figure}
        \centering
        \includegraphics[width=\textwidth]{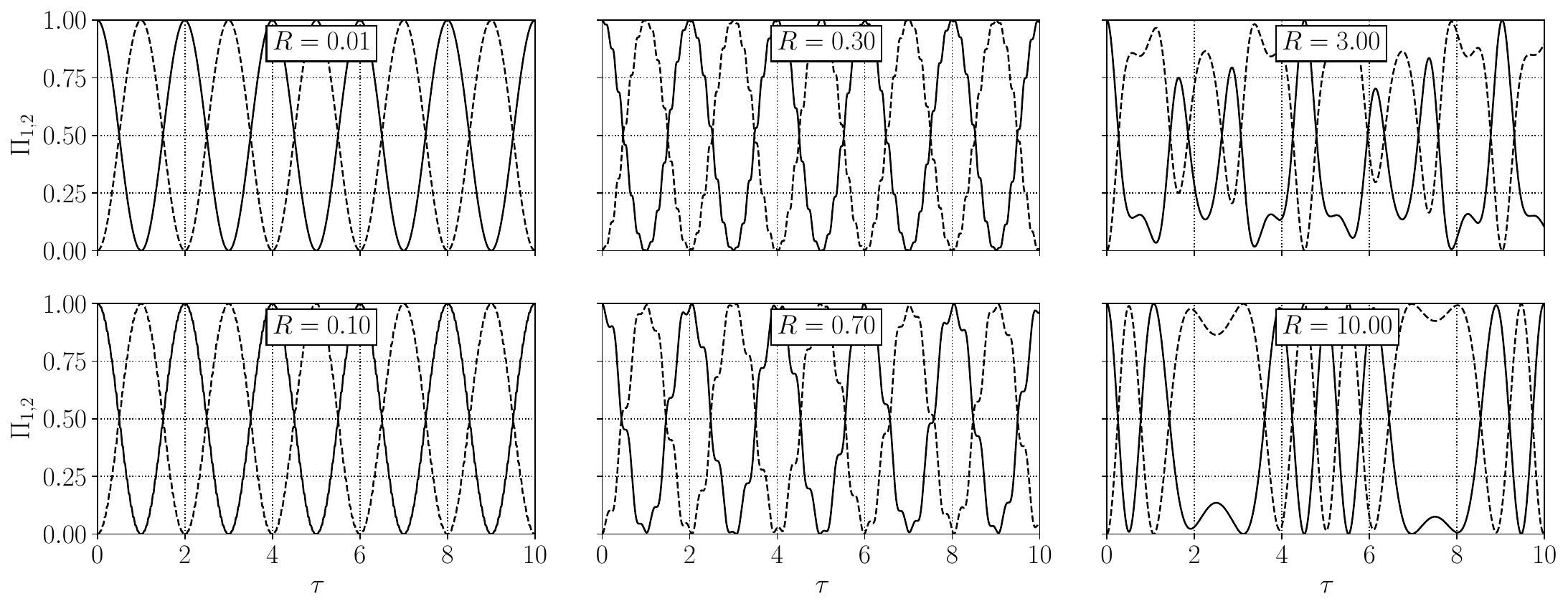}
        \caption{Numerical solution for population dynamics of a two-level system driven at resonance.
        Horizontal axis is in units of proper time $\tau$.
        Each subplot corresponds to a different normalized effective drive intensity $R$, indicated on the plot and corresponding to respective plot in Fig.~\ref{fig:RabiProfile_Multi}. 
        The full and dotted lines show level populations $\Pi_1(t)$ and $\Pi_2(t)$, respectively.}
        \label{fig:TwoLevel_Multi}
    \end{figure}  
    
    \end{widetext}  
    
\subsubsection{Off-Resonant Drive Dynamics}

\begin{figure}
    \centering
    \includegraphics[width=\columnwidth]{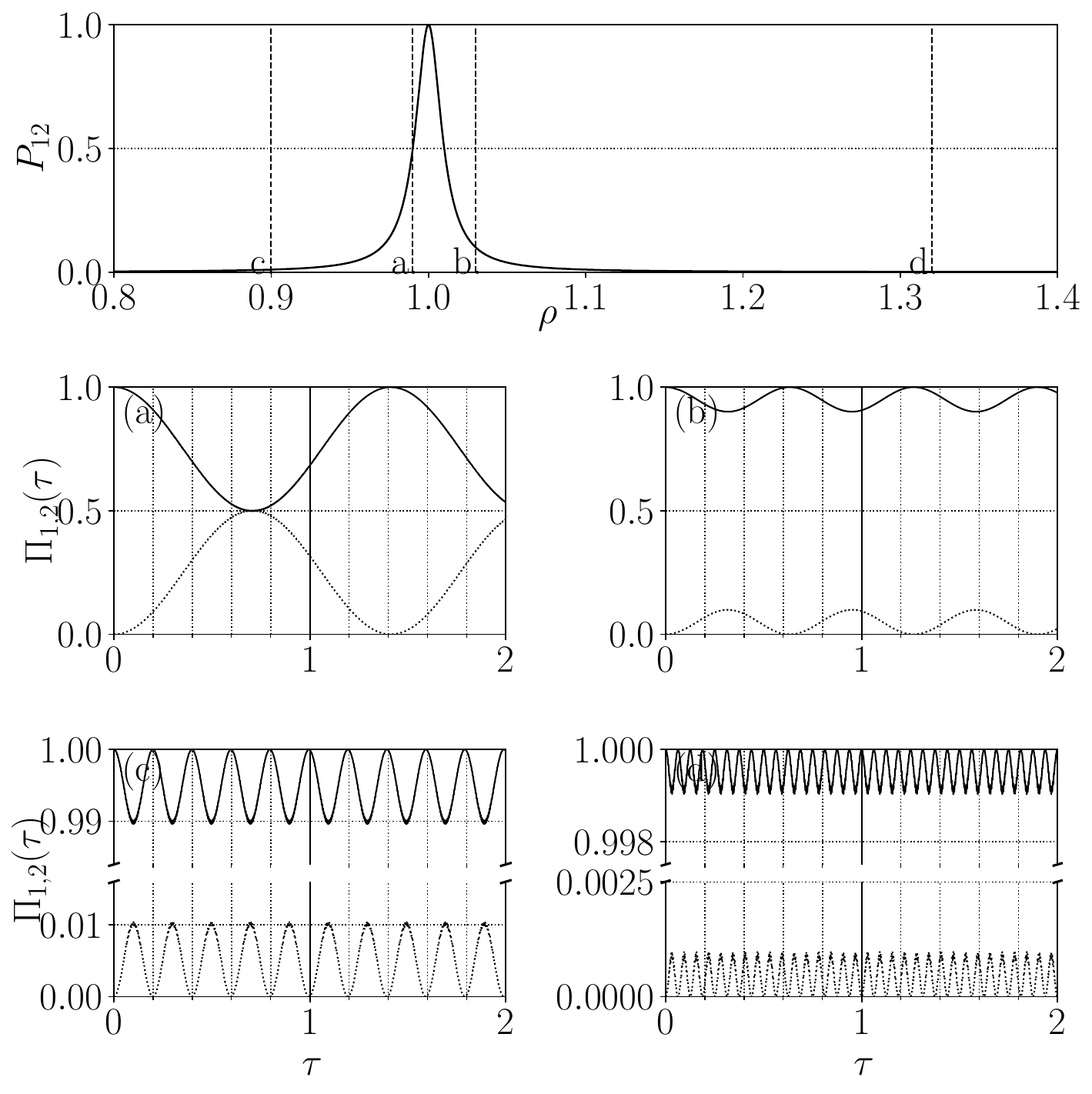}
    \caption{Population dynamics for off-resonant drive.
    Top-row plot presents the Rabi profile for the system transition with $R = 0.01$, well within the RWA validity range. Vertical lines indicate the four off-resonant drive frequencies used: (a) $\rho = 0.99 / P = 0.5$; (b) $\rho = 1.03 / P = 0.1$; (c) $\rho = 0.9 / P = 0.01$; (d) $\rho = 1.32 / P = 0.001$.
    The four bottom plots display numerical simulations of dynamics of the system for the four selected drive frequencies.
    The range of the horizontal axis is set to the value of $\tau = 2$ which corresponds to one full population oscillation in resonant case.}
    \label{fig:off-resonance}
    \end{figure}

    To illustrate the validity of the RWA solution in the case of off-resonant driving, Figure~\ref{fig:off-resonance} shows the same two-level system driven at four different off-resonant frequencies. 
    We assume that the condition for a well behaved RWA solution, Eq.~\eqref{eq:RWA-robust-threshold} is fulfilled---indeed, we set $R = 0.01$---so that the system behaves predictably at any of the drive frequencies of the displayed range.
    The numerical solutions closely follow the RWA predictions in all cases: 
    as detuning increases, the amplitude of the oscillations reduces whereas their frequency increases.
    
\subsection{Many-level systems: Rabi leak and Rabi spectrum}
\label{many level}

    So far, we considered the Rabi profile in the context of an isolated two-level system.
    In realistic physical situation, each of the two targeted levels is typically coupled to multiple other levels. 
    In these settings, quantum control---maximizing the efficiency of the targeted transition---requires minimizing unwanted population transfer---i.e. transfer leakage---into these additional levels.

\subsubsection{Rabi Leak}

    In many-level quantum systems, the main obstacle to clean population transfer along a chosen transition is not the impact of counter-rotating terms, but the unwanted excitation of neighboring transitions.
    
    These transfer leakage effects appear at drive intensities far below the ones that yield Bloch-Siegert oscillations, when the drive intensity is still well within the limits set by Eq.~\eqref{eq:simplified-RWA-validity-criterion}. 
    In that case, Rabi profiles of all the relevant transitions provide an excellent description of the actual system dynamics.
    
    For a targeted transition $|i\rangle \!\leftrightarrow\! |j\rangle$, we define the \emph{Rabi leak} to an untargeted transition $|i\rangle \!\leftrightarrow\! |n\rangle$ ($n\neq j$) as the value of the Rabi profile of the leak transition evaluated at the drive frequency resonant with the targeted one:
    \begin{equation}
    \label{eq:leakprofile}
        L_{ij}^{n}(F)
        := P_{in}(F, \omega_{ij})
        = \frac{1}{1 + \left( \frac{\hbar(\omega_{ij} - \omega_{in})}{F I_{in}} \right)^2 } ,
    \end{equation}
    where $I_{in}$ and $\omega_{in}$ are the coupling coefficient and resonant frequency of the leak transition.  
    Thus, $L_{ij}^{n}$ is simply the height of the untargeted Lorentzian (its Rabi profile) at the frequency of the targeted one.
    
    The quantity $L_{ij}^{n}(F)$ represents the maximum fractional population that can leak from $|i\rangle$ into $|n\rangle$ while the system is driven resonantly on $|i\rangle \!\leftrightarrow\! |j\rangle$. 
    Analogous expressions define the leaks from $|j\rangle$ to other levels $|m\rangle$ ($m\neq i$).
    
    For small leakage amplitudes ($L_{ij}^{n}\!\ll\!1$), Eq.~\eqref{eq:leakprofile} can be expanded to the leading order in $F$:

    \begin{equation}
        L_{ij}^{n}(F) 
        \approx 
        \frac{(I_{in})^2}{\hbar^2(\omega_{ij} - \omega_{in})^2} F^2 
          + \mathcal{O} (F^4).
    \end{equation}
    Summing over all possible leakage paths yields the total leak for the target transition:
    \begin{widetext}
    \begin{equation}
    \label{eq:total-leak}
        L_{ij}(F)
        :=
        \sum_{n \neq j} L_{ij}^{n}(F)
        + \sum_{m \neq i} L_{ij}^{m}(F)        
        \approx
        F^2
        \!\left(
            \sum_{n \neq j} \frac{I_{in}^2}{\hbar^2(\omega_{ij} - \omega_{in})^2}
            + \sum_{m \neq i} \frac{I_{jm}^2}{\hbar^2(\omega_{ij} - \omega_{jm})^2}
        \right).
    \end{equation}
    \end{widetext}
    Inverting Eq.~\eqref{eq:total-leak} gives the maximum drive amplitude $F_{\max}$ consistent with a tolerable leak $L_{\max}\!\ll\!1$:
    \begin{equation}
    \label{eq:f0max}
        F_{\max} \approx
        \sqrt{
            \frac{L_{\max}}{
                \displaystyle
                \sum_{n \neq j} 
                    \frac{I_{in}^2}{\hbar^2(\omega_{ij} - \omega_{in})^2}
                + \sum_{m \neq i}
                    \frac{I_{jm}^2}{\hbar^2(\omega_{ij} - \omega_{jm})^2}
            } }.
    \end{equation}
        
    When leakage is dominated by a single untargeted transition, this simplifies to:
    \begin{equation}
    \label{eq:F_max-single}
        F_{\max}
        \approx 
        \frac{\hbar \, |\omega_{\text{target}} - \omega_{\text{leak}}|}
             {|I_{\text{leak}}|}
        \sqrt{L_{\max}},
    \end{equation}
    where $\omega_{\text{target}}$ and $\omega_{\text{leak}}$ are the (angular) transition frequencies of the targeted and leak transitions, respectively, and $I_{\text{leak}}$ is the coupling coefficient of the leak transition.
    
    From Eq.~\eqref{tau} we can obtain the minimum laboratory time required to achieve a single complete population inversion ("$\pi$-pulse time") on the targeted transition while keeping leakage below $L_{\max}$:
    \begin{equation}
    \label{eq:min-transfer-time}
        t_{\min}^{\pi} = \frac{2\pi \hbar}{F_{\max}\, I_{\text{target}}}.
    \end{equation}
    Here, $I_{\text{target}}$ is the coupling coefficient for the target
    transition.
    
    Again, for the single dominant leak case, substituting Eq.~\eqref{eq:F_max-single} into Eq.~\eqref{eq:min-transfer-time} yields:
    \begin{equation}
    \label{eq:min-transfer-time-single}
        t_{\min}^{\pi}
            =
            \frac{1}{|f_{\text{target}} - f_{\text{leak}}|}
            \frac{|I_{\text{leak}}|}{|I_{\text{target}}|}
            \frac{1}{\sqrt{L_{\max}}},
    \end{equation}
    where $f_{\text{target}}=\omega_{\text{target}}/2\pi$ and $f_{\text{leak}}=\omega_{\text{leak}}/2\pi$ are the corresponding (plain) transition frequencies in hertz.
	
\subsubsection{Rabi spectrum}    
    Rabi leaks related to a particular targeted transition can be visualized using \textit{Rabi spectrum} of that transition: the collection of all relevant Rabi profiles---targeted and leakage ones---for that transition, plotted against a common frequency axis. Such representation immediately reveals which off-resonant transitions are likely to contribute most to leakage.
    
\begin{figure}
        \centering
        \includegraphics[width=\columnwidth, trim=0 20 60 0, clip]{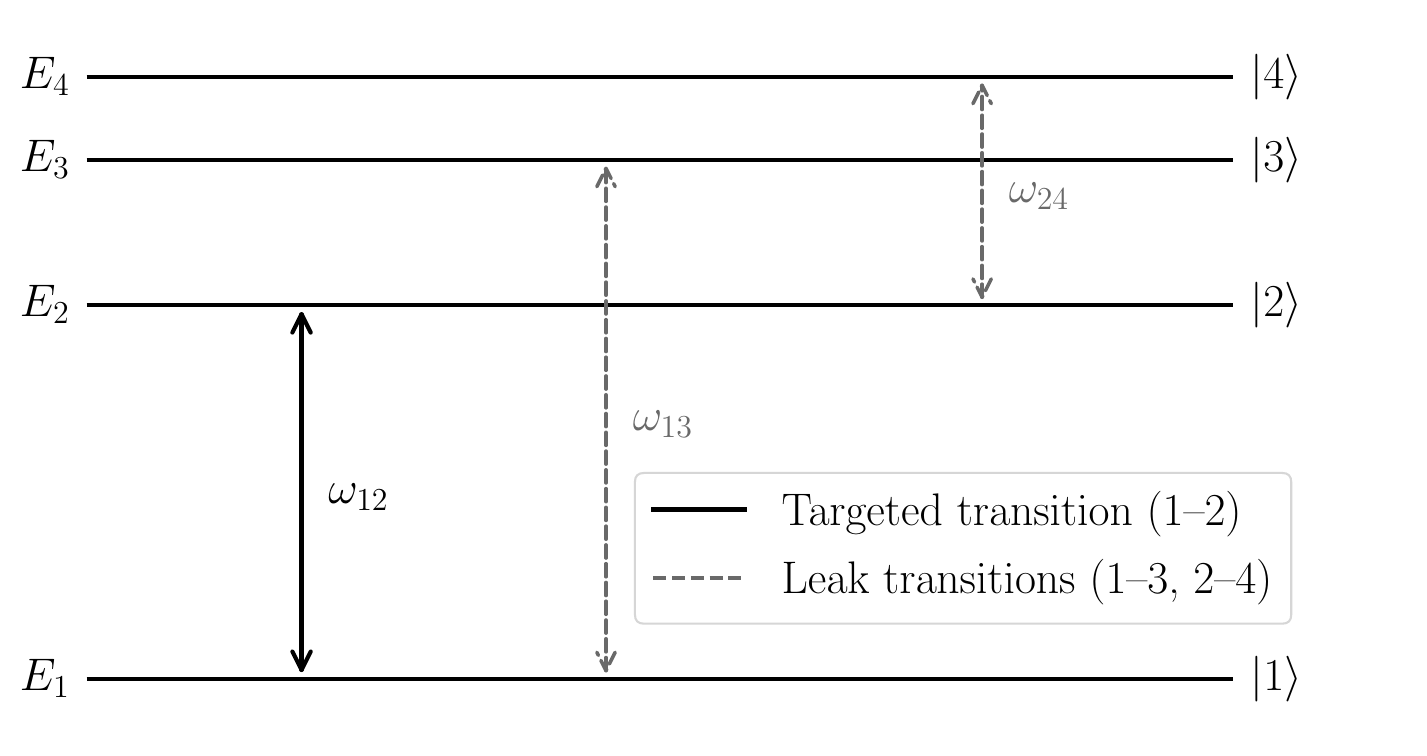}
        \caption{Energy diagram illustrating a targeted transition and leakage channels in a four-level externally driven system. The solid black double arrow denotes the driven transition $\ket{1} \leftrightarrow \ket{2}$ at frequency $\omega_{12}$. The dashed gray arrows indicate leakage transitions $\ket{1} \leftrightarrow \ket{3}$ ($\omega_{13}$) and $\ket{2} \leftrightarrow \ket{4}$ ($\omega_{24}$). In this configuration the targeted frequency lies between the two leakage frequencies, $\omega_{24} < \omega_{12} < \omega_{13}$ .}
        \label{fig:energy-levels}
    \end{figure}

\begin{widetext}        
    
    \begin{figure}
        \centering
        \includegraphics[width=\textwidth]{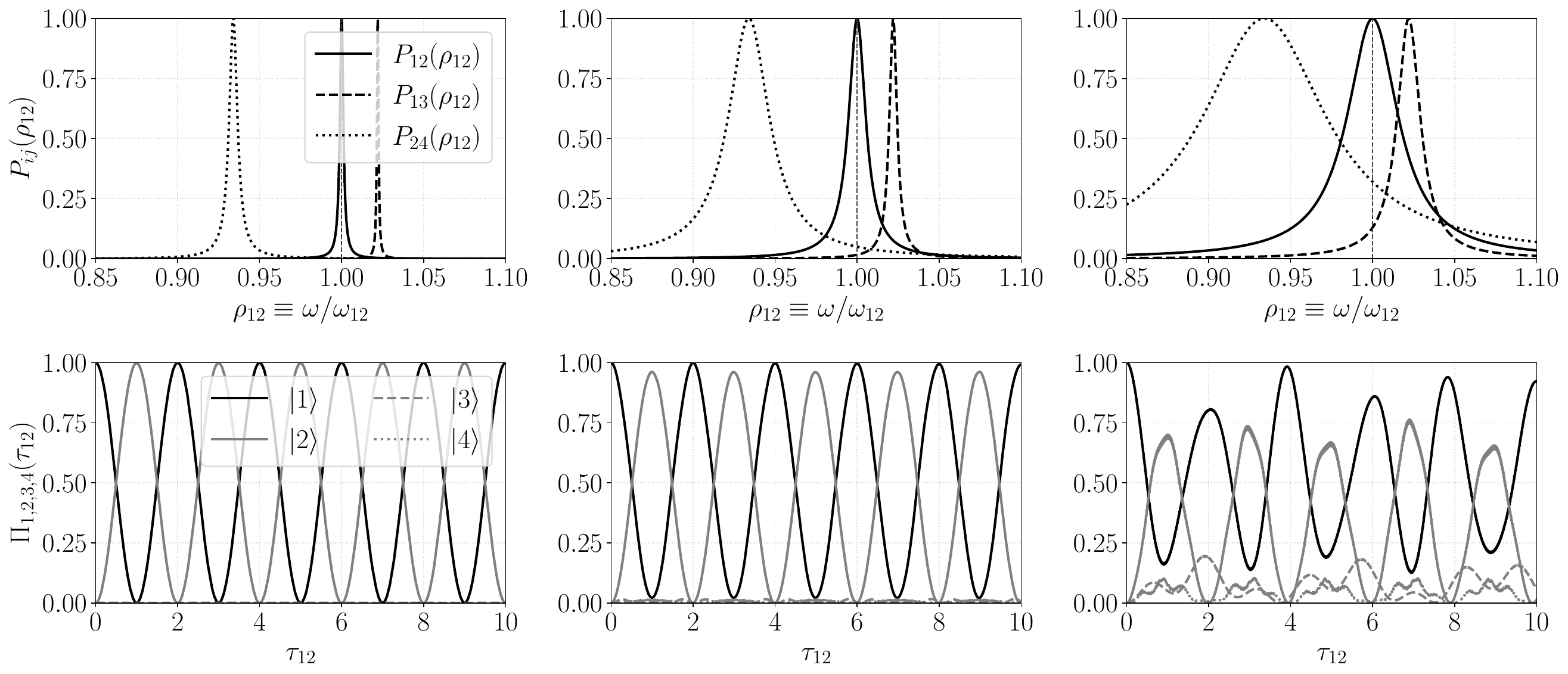}
        \caption{Rabi spectrum (top row) and corresponding numerically simulated, full lab-frame dynamics (bottom row) of the four-level system from Fig.\ref{fig:energy-levels}, driven at the resonant frequency of the transition $|1\rangle \leftrightarrow |2\rangle$, $\omega_{12}$.
        Each row corresponds to a different effective drive amplitude. The x-axis of the plot is normalized with respect to the targeted transition, so that $\rho_{12} = \tfrac{\omega}{\omega_{12}}$.}
        \label{fig:rabi-spectrum}
    \end{figure}    
    
    \end{widetext}

    For example, consider a four-level system consisting of states $|i\rangle, \, i=1,2,3,4,$ where the drive couples $|1\rangle \leftrightarrow |2\rangle$, $|1\rangle \leftrightarrow |3\rangle$ and $|2\rangle \leftrightarrow |4\rangle$ (Fig.\ref{fig:energy-levels}). Rabi spectrum for the transition $|1\rangle \leftrightarrow |2\rangle$ is simply the superposition of $P_{12}(\omega)$, $P_{13}(\omega)$ and $P_{24}(\omega)$ (top row plots in Fig.\ref{fig:rabi-spectrum}). 
        
    The Rabi spectrum makes it clear that although the transition $|1\rangle \leftrightarrow |3\rangle$ is spectrally closer to the target $|1\rangle \leftrightarrow |2\rangle$ than is $|2\rangle \leftrightarrow |4\rangle$, since the coupling for $|2\rangle \leftrightarrow |4\rangle$ is stronger than the coupling for $|1\rangle \leftrightarrow |3\rangle$ the main contribution to leakage should come from $|2\rangle \leftrightarrow |4\rangle$.
    
    The three bottom row plots of Fig.\ref{fig:rabi-spectrum} display the actual (numerically simulated) time-domain dynamics of the same system, for three different effective drive amplitudes. Observe that the perturbation of the leakage level $|3\rangle$ dominates, seemingly contrary to the prediction of the Rabi spectrum. This is due to the fact that the system starts in state $|1\rangle$, which is coupled only to $|2\rangle$ and $|3\rangle$, but not directly to $|4\rangle$. So, leakage to $|4\rangle$ begins only after a portion of the population has been transferred to $|2\rangle$, which is coupled to $|4\rangle$. Nevertheless, the Rabi spectrum correctly predicts the magnitude of the total leakage, thus efficiently charting the limits of clean population transfer via targeted transition.    
    
    \subsection{Illustrative Example from Real Systems - Transmon Qubits}
    \label{sec:Appendix}
    
    To demonstrate the practical utility of the Rabi profile framework, and in particular of concepts of Rabi leak and Rabi spectrum, we now consider one representative physical system where controlled coherent driving of quantum transitions is of central importance.
    
    \subsubsection{Superconducting Transmon Qubits}
    
    Transmon qubits are weakly anharmonic superconducting oscillators derived from the Cooper-pair box, engineered to suppress charge noise by operating at large $E_J/E_C$ ratios~\cite{koch2007,blais2021,wang2022npj}.
    They form the backbone of current circuit architectures for quantum computing, offering long coherence times, fast gate speeds, and integration with microwave control hardware~\cite{krantz2019}. 
    
    The computational subspace that plays the role of a qubit in a transmon is spanned by the ground state and the first excited state, so the relevant target transition for qubit operations is $|0\rangle \leftrightarrow |1\rangle$.
    The nearly harmonic spectrum of a transmon poses a challenge for fast and selective qubit operations, making leakage out of the computational subspace to higher levels---in particular from $|1\rangle|$ to $|2\rangle$---the primary error mechanism in fast driving.
    
    The Rabi leak formula enables precise quantification of the minimum allowable population transfer time, while the complementary Rabi spectrum plot provides a direct way to visualize this selectivity limit. 
    This, in turn, sets the upper limit on the speed of the computing operations of a transmon-based computer.
    
    \subsubsection{Theoretical Predictions Using the Rabi Profile Framework}
    The operation of a transmon qubit thus depends on two transitions: the targeted $|0\rangle \!\leftrightarrow\! |1\rangle$ transition and the dominant leakage channel $|1\rangle \!\leftrightarrow\! |2\rangle$.  
    This situation corresponds exactly to the single-leak case described by Eq.~\eqref{eq:min-transfer-time-single}.
    
    In the transmon, the coupling coefficients $I_{ij}$ are given by the charge–matrix elements of the Cooper-pair number operator,  $n_{ij}=\langle i|\hat{n}|j\rangle$, which determine the electric-dipole strength of each transition.  
    These matrix elements satisfy the approximate relation~\cite{koch2007}:
    \begin{equation}
    	\label{eq:transmon-relation}
    	\frac{I_{12}}{I_{01}}
    	\equiv
    	\frac{n_{12}}{n_{01}}
    	\approx
    	\sqrt{2}.
    \end{equation}
    Substituting this relation into Eq.~\eqref{eq:min-transfer-time-single} yields
    \begin{equation}
    	\label{eq:tpi_min-transmon}
    	t_{\pi}^{\min}
    	= \frac{\sqrt{2}}{|f_{01}-f_{12}|}\,
    	\frac{1}{\sqrt{L_{\max}}}.
    \end{equation}
    
    Inserting into this expression the experimental values of the two 3D transmon transition frequencies measured by Peterer \emph{et al.}~\cite{peterer2015}:
    \begin{equation}
    	f_{01}=4.97~\mathrm{GHz}, \qquad
    	f_{12}=4.70~\mathrm{GHz},
    \end{equation}
    yields the minimum gate duration permitted for a specified leakage tolerance $L_{\max}$.  
    This quantity provides a directly measurable, system-independent lower bound on transmon gate times, linking the observed transition frequencies to the ultimate spectral limit of coherent control.,
    
    Table~\ref{tab:transmon_times} summarizes the resulting minimum $\pi$-pulse times for three specific values of leakage tolerance: $0.1\%$, $1\%$ and $10\%$.
    
    \begin{table}[t]
    	\centering
    	\caption{Minimum $\pi$-pulse times $t_{\min}^\pi$ for a transmon qubit, evaluated for different leakage tolerances $L_\text{max}$. Values are calculated using Eq.~\eqref{eq:tpi_min-transmon} with experimental parameters from~\cite{peterer2015}.}
    	\begin{tabular}{c c c}
    		\hline\hline
    		$L_\text{max}$ \qquad & $t_{\min}^\pi$ (ns) \\
    		\hline
    		0.1\% & 165.56 \\
    		1\%   & 52.38 \\
    		10\%  & 16.55  \\
    		\hline\hline
    	\end{tabular}
    	\label{tab:transmon_times}
    \end{table}
    
\begin{figure}
    	\centering
    	\includegraphics[width=8.5cm]{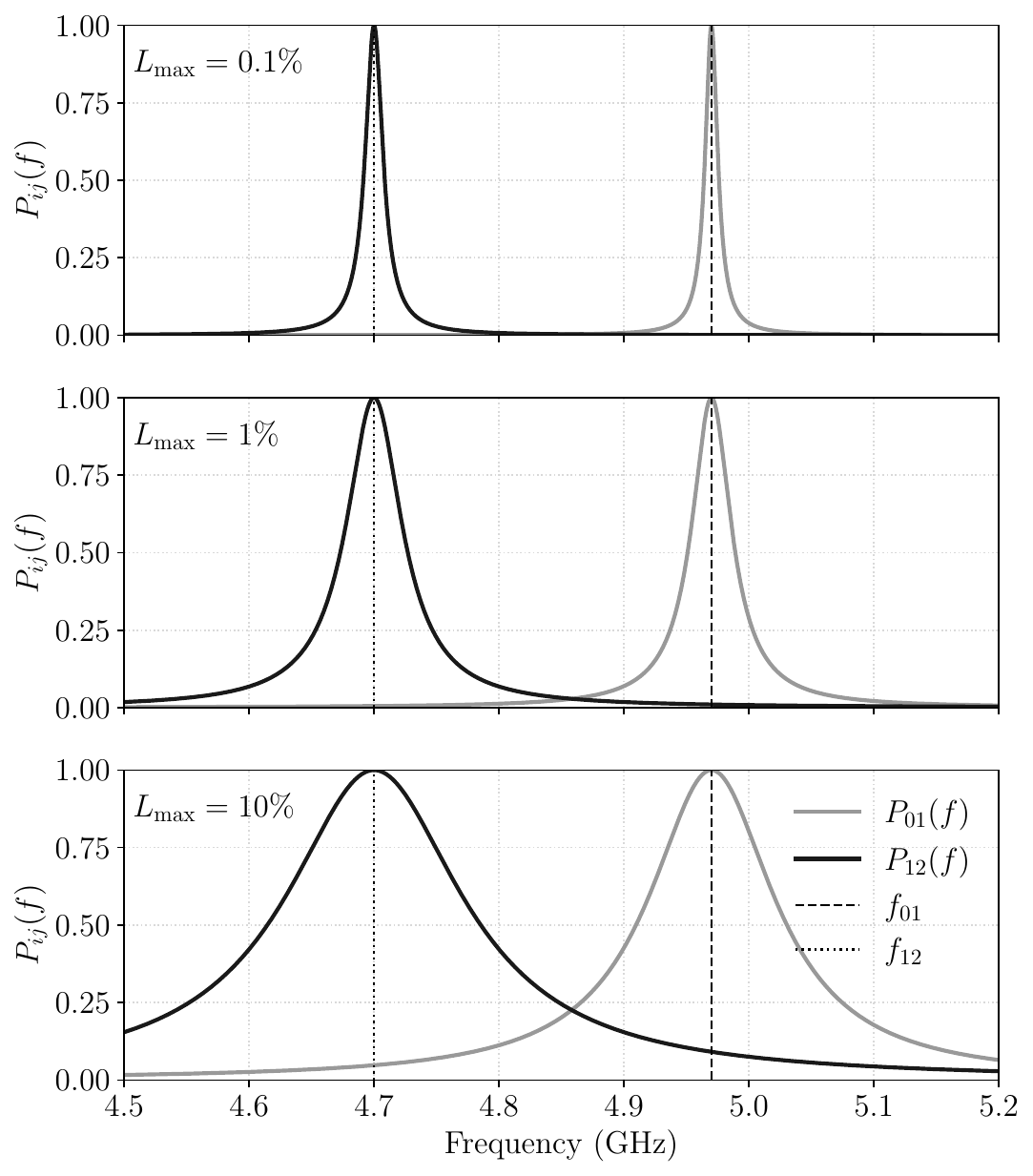}
    	\caption{Rabi spectra for a superconducting transmon qubit with parameters from~\cite{peterer2015} and Table~\ref{tab:transmon_times}. The panels show Lorentzian excitation profiles for the $|0\rangle \! \leftrightarrow\!|1\rangle$ (light gray) and $|1\rangle \! \leftrightarrow\! |2\rangle$ (dark gray) transitions, with vertical dashed and dotted lines marking their resonant frequencies $f_{01}$ and $f_{12}$, respectively. The linewidths are set by the leak-limited drive $F_{\max}(L_{\max})$, with tolerated leakage $L_{\max}$ of 0.1\% (top), 1\% (middle), and 10\% (bottom). Increasing $L_{\max}$ broadens the profiles, allowing faster $\pi$-pulses at the cost of greater off-resonant excitation.}
    	\label{fig:transmon_spectrum}
    \end{figure}

    Figure~\ref{fig:transmon_spectrum} illustrates the Rabi spectrum for the $|0\rangle \leftrightarrow |1\rangle$ and $|1\rangle \leftrightarrow |2\rangle$ transitions using the drive amplitude parameters from Table~I. 
    The broadening and eventual overlap of the $P_{12}(\omega)$ profile with $\omega_{01}$ directly signals the onset of leakage. The greater the overlap, the greater the expected leakage.
    
    \subsubsection{Agreement With Experimental Data}
    Our predictions in Table~\ref{tab:transmon_times} indicate that transmon qubit gates cannot be coherently driven with durations shorter than roughly 10–20~ns, since such times would correspond to extreme leakage comparable in magnitude to the amplitude of transfer of the targeted population.  
    This estimate is consistent with experimental observations of gate performance in superconducting transmons.
    
    In large-scale superconducting processors, typical single-qubit gate times fall in the range of 20–100~ns, with faster operations generally accompanied by increased leakage and error rates.  
    For example, Barends \emph{et al.} demonstrated high-fidelity gates at the surface-code threshold with $\pi$-pulse times around 20–40~ns~\cite{barends2014}, while Kelly \emph{et al.} used optimal-control techniques to suppress leakage and improve fidelities within the same regime~\cite{kelly2014}.  
    More recently, Werninghaus \emph{et al.} achieved 4.16~ns single-qubit pulses with 99.76\% fidelity and only 0.044\% leakage~\cite{werninghaus2021}, among the fastest reported to date.
    
    There is, however, one major difference between our predictions and these experimental results: 
    the usual leak rates in all these experiments are around $\sim 10^{-5}$--$10^{-4}$~\cite{wood2018}, which is orders of magnitude lower than our nominal 10\%. 
    The apparent discrepancy arises from the different nature of the driving fields. 
    Experimental results rely on carefully shaped pulses with fully optimized duration and frequency spectrum, whereas our Rabi profile framework assumes a simple, perfectly sinusoidal resonant drive, presenting the broadest possible spectrum for a given duration and thus the least favorable case for spectral selectivity.
    
    As analytically demonstrated by Bonacci \textit{et al.}~\cite{bonacci2003.2,bonacci2005}, just replacing a pure flat sinusoidal drive with a pulse of the same maximum amplitude but optimized envelope and frequency chirp can reduce a 10\% leak to negligible levels. 
    Further refinements using established techniques such as GRAPE~\cite{kelly2014} and DRAG~\cite{motzoi2009}, as well as more recent analytically motivated envelope designs~\cite{hyyppa2024}, have demonstrated comparable benefits through hybrid analytic–numerical optimization.
    
    However, such optimization remains effective only as long as the interference from the neighboring transition remains "perturbative". 
    The Rabi profile framework precisely quantifies where this boundary lies and thereby defines a universal crossover between regimes of controllability:
    \begin{itemize}
    	\item for Rabi leak values below approximately 10\%, the influence of the leak transition can be efficiently mitigated by pulse shaping, phase compensation, or analytical-envelope optimization; 
    	\item once the predicted Rabi leak increases beyond roughly 30\%, the spectral overlap between transitions becomes intrinsic to the system, and no pulse-engineering method can fully compensate for it. 
    \end{itemize}
    In this sense, the Rabi profile approach identifies the fundamental limit separating the \emph{fine-tuning regime}---where experimental ingenuity can still improve coherence---from the \emph{spectral-overlap regime}, where the hardware itself imposes a hard lower bound on the speed of coherent control.
    
    \subsubsection{Practical Application: Optimized Pulse Design}
    The “10\% leak cutoff” criterion provides a practical foundation for applying the Rabi profile framework to the design of real quantum hardware.
    By fixing key parameters in Eq.~\eqref{eq:min-transfer-time-single} as follows:
    \begin{itemize}
    	\item setting $L_\text{max}=10\%$ as the maximum leakage that can still be fully compensated by pulse shaping, and
    	\item assuming that in most physical systems the coupling strengths for the relevant transitions are comparable, i.e. such that $\frac{|I_\text{leak}|}{|I_\text{target}|} \approx 1$,
    \end{itemize}
    we obtain a remarkably simple, system-independent design rule:
    \begin{equation}
    	\label{eq:tpi_design_rule}
    	t_\pi^\text{min}\,[\mathrm{ns}]
    	\;\approx\;
    	\frac{3.16}{|f_{\text{target}}-f_{\text{leak}}|\,[\mathrm{GHz}]}\,.
    \end{equation}
    
    Eq.~\eqref{eq:tpi_design_rule} directly links the shortest achievable coherent $\pi$-pulse duration to experimentally measurable spectroscopic quantities—specifically, the frequency separation between the targeted transition and its nearest leakage transition.  
    It therefore provides an intuitive bridge between spectroscopy and hardware design: for a given level spacing, it sets the absolute lower bound on coherent operation speed before off-resonant excitation becomes non-remediable.  
    
    Using Eq.~\eqref{eq:tpi_design_rule}, we can estimate the minimal achievable $\pi$-pulse durations that produce coherent population transfer across representative qubit technologies.  
    Table~\ref{tab:design_times} summarizes these estimates, based on typical transition separations $\Delta f = |f_{\text{target}}-f_{\text{leak}}|$ reported in the literature.  
    As expected, platforms with larger intrinsic anharmonicity (larger $\Delta f$) permit faster coherent gates before leakage becomes prohibitive, while nearly harmonic systems are fundamentally slower, regardless of control optimization.
    
    \begin{table}[h!]
    	\centering
    	\caption{Estimated lower bounds for coherent $\pi$-pulse durations ($L_{\max}=10\%$, $|I_{\text{leak}}|/|I_{\text{target}}|\!\approx\!1$), using Eq.~\eqref{eq:tpi_design_rule}. References provide typical values of the neighboring transition spacing $\Delta f$.}
    	\begin{tabular}{lcc}
    		\hline\hline
    		\textbf{Qubit platform} & $\Delta f$ (GHz) & $t_\pi^\text{min}$ (ns) \\
    		\hline
    		3D/planar transmon~\cite{peterer2015, burnett2019, werninghaus2021, wang2025} & 0.1--0.3 & 10--30 \\
    		Fluxonium/CSFQ~\cite{manucharyan2009,nguyen2019,moskalenko2022,najera2024,earnest2018,ku2020} & 0.6--4 & 0.8--5 \\
    		NV center (electron spin)~\cite{childress2006,dolde2014} & 0.5--3 & 1--6 \\
    		\hline\hline
    	\end{tabular}
    	\label{tab:design_times}
    \end{table}
    
    Based on this table, the Rabi profile framework suggests a simple, system independent, hierarchical approach to pulse design and device optimization for qubit technologies:
    \begin{enumerate}
    	\item \textbf{Select a physical platform.}  
    	Use Table~\ref{tab:design_times} to identify a system whose characteristic minimum transfer time $t_\pi^\text{min}$ aligns with the desired gate speed range.
    	\item \textbf{Determine the corresponding drive amplitude.}  
    	From Eq.~\eqref{eq:f0max} or its single-leak simplification~\eqref{eq:F_max-single}, compute the leak-limited drive $F_{\max}$ for your chosen system.  
    	Use this as the baseline amplitude for pulse construction.
    	\item \textbf{Draft the analytically optimized initial pulse shape.}  
    	Employ closed-form pulse-design techniques introduced in~\cite{bonacci2003.2,bonacci2005} to generate an initial chirped waveform with the same $F_{\max}$, minimizing leakage analytically without numerical optimization.
    	\item \textbf{Numerically fine-tune the pulse.}  
    	Apply standard numerical optimization methods such as DRAG or GRAPE to further suppress residual leakage, improve phase fidelity, and adapt the waveform to hardware-specific bandwidth constraints.
    \end{enumerate}
    
    This protocol provides a \emph{system-independent} recipe for approaching the physical limits of quantum control: 
    the analytical Rabi profile framework establishes the fundamental time–frequency bound;  
    the earlier analytical results reach close to this bound through deterministic pulse design;  
    and modern numerical techniques refine the solution within the same physical constraints.
    
    \subsubsection{Broader Applications Beyond Quantum Computing}
    While we have focused on transmon qubits as a specific example, it is worth emphasizing that the relationship between system parameters and minimal transfer times derived here is far more general. 
    This framework can be applied to a variety of contexts where coherent control over state transitions is required.
    
    In atomic and molecular physics, for instance, the same principles can guide the design of laser pulses to drive specific electronic transitions while minimizing unintended excitations of neighboring states. 
    This is particularly valuable in precision spectroscopy and controlled chemical reactions.
    
    Similarly, in nuclear magnetic resonance (NMR) and magnetic resonance imaging (MRI), the same approach can help design pulse sequences that selectively excite or manipulate certain spin transitions without affecting others, optimizing the fidelity of NMR or MRI procedures. 
    
    Finally, even in classical systems such as coupled mechanical or optical oscillators, analogous principles can be used to achieve selective energy transfer with minimal unintended excitation.
    
    In conclusion, this system-independent strategy offers a roadmap for applying the Rabi profile framework not only to transmon qubits but to a wide range of quantum and even classical resonant systems, providing a universal design guideline for achieving minimal-time coherent control.

\section{Conclusion}
        
    Although the underlying derivation is straightforward, the Rabi profile framework provides immediate practical insight.  
    It enables quick estimates of maximum usable drive strengths, minimum transfer times, and other design-relevant constraints imposed by the system’s internal structure.  
    Once transition frequencies and dipole couplings are known, the Rabi leak and spectrum of all allowed transitions can be readily computed, revealing which channels permit high-fidelity control under a given drive amplitude.  
    
    Our analysis is limited to coherent dynamics. In real systems, spontaneous decay and other dissipative channels are also present. 
    However, their main effect is to broaden the resonance profiles and reduce the maximum achievable transfer amplitudes, while leaving the graphical construction of Rabi profiles unchanged.
    In practice, spontaneous decay can be included as an additional damping channel, treated independently of the coherent drive considered here \cite{merlin2021, cohentannoudji1992,scully1997}. 
    
    The conceptual simplicity of this approach makes it equally valuable in pedagogy and research.  
    It offers an intuitive  frequency-domain visualization of driven quantum dynamics that complements more formal control theory.  
    By revealing how spectral separability governs the speed and fidelity of population transfer, the Rabi profile framework establishes a unifying perspective that connects quantum control, system design, and information-theoretic limits within a single, accessible framework.

\end{document}